\documentclass[preprint]{elsarticle}

\usepackage[hidelinks]{hyperref}

\usepackage[utf8]{inputenc}
\usepackage[T1]{fontenc}

\usepackage{amsmath,amsthm,amssymb}
\usepackage{mathtools}
\usepackage{subcaption}

\usepackage{tikz}
\usepackage{pgf-pie}
\usepackage{pgfplots}
\usepackage{xcolor} %needed to define colors using hex code in tikz 
\definecolor{bluepie1}{HTML}{ece7f2}
\definecolor{bluepie2}{HTML}{a6bddb}
\definecolor{bluepie3}{HTML}{2b8cbe}
\usepackage{environ} %scale tikzpicture to textwidth
\makeatletter
\newsavebox{\measure@tikzpicture}
\NewEnviron{scaletikzpicturetowidth}[1]{%
	\def\tikz@width{#1}%
	\def\tikzscale{1}\begin{lrbox}{\measure@tikzpicture}%
		\BODY
	\end{lrbox}%
	\pgfmathparse{#1/\wd\measure@tikzpicture}%
	\edef\tikzscale{\pgfmathresult}%
	\BODY
}
\makeatother
\usetikzlibrary{shapes}

\usepackage{tabularx,booktabs,ctable}
\usepackage{threeparttable}
\usepackage{siunitx} % align columns at decimal point in table
\usepackage{xltabular} % tabularx mit Seitenumbruch!
\usepackage{caption}
\usepackage{float}
\usepackage{rotating}
\newcommand\tabrotate[1]{\begin{turn}{20}\rlap{#1}\end{turn}}
\usepackage{multirow}

\usepackage[markup=nocolor]{changes}
\usepackage{textcomp}
\usepackage{enumitem}
\usepackage{todonotes}
\usepackage{subcaption}

\usepackage{natbib}
\biboptions{round,authoryear}

\theoremstyle{definition}

\theoremstyle{remark}

\usepackage{arydshln}
\makeatletter
\def\adl@drawiv#1#2#3{%
	\hskip.5\tabcolsep
	\xleaders#3{#2.5\@tempdimb #1{1}#2.5\@tempdimb}%
	#2\z@ plus1fil minus1fil\relax
	\hskip.5\tabcolsep}
\newcommand{\cdashlinelr}[1]{%
	\noalign{\vskip\aboverulesep
		\global\let\@dashdrawstore\adl@draw
		\global\let\adl@draw\adl@drawiv}
	\cdashline{#1}
	\noalign{\global\let\adl@draw\@dashdrawstore
		\vskip\belowrulesep}}
\makeatother

\usepackage{xspace}
\newcommand{\hma}{\textsl{\textsf{Hol~mich!~App}}\xspace}

\usepackage{chngcntr}

\begin{document}
	\begin{frontmatter}
		 \author[1]{Hayk Asatryan}
		\ead{asatrianh@gmail.com}
		\author[2]{Daniela Gaul\fnref{fn2}}
		\ead{gaul@math.uni-wuppertal.de}
		\fntext[fn2]{Corresponding author.}
		\author[3]{Hanno Gottschalk}
		\ead{gottschalk@math.tu-berlin.de}
		\author[2]{Kathrin Klamroth}
		\ead{klamroth@uni-wuppertal.de}
		\author[2]{Michael~Stiglmayr}
		\ead{stiglmayr@uni-wuppertal.de}
		\title{Ridepooling and public bus services: A comparative case-study}
		\address[1]{Bochum University of Applied Sciences, 44801 Bochum, Germany}
		\address[2]{School of Mathematics and Natural Sciences, University of Wuppertal, 42119 Wuppertal, Germany}
		\address[3]{Faculty II - Mathematics and Natural Sciences, TU Berlin, 10623 Berlin, Germany}
		
		\journal{EURO Journal on Transportation and Logistics}

\begin{abstract} 
This case-study aims at a comparison of the service quality of time-tabled buses as compared to on-demand ridepooling cabs in the late evening hours in the city of Wuppertal, Germany. To evaluate the service quality of ridepooling as compared to bus services, and to simulate bus rides during the evening hours, transport requests are generated using a predictive simulation. To this end, a framework in the programming language R is created, which automatically combines generalized linear models for count regression to model the demand at each bus stop. Furthermore, we use classification models for the prediction of trip destinations. To solve the resulting dynamic dial-a-ride problem, a rolling-horizon algorithm based on the iterative solution of Mixed-Integer Linear Programming Models (MILP) is used. A feasible-path heuristic is used to enhance the performance of the algorithm in presence of high request densities. This allows an estimation of the number of cabs needed depending on the weekday to realize the same or a better general service quality as the bus system.
\end{abstract}

\end{frontmatter}

\section{Introduction}\label{sec:intro}

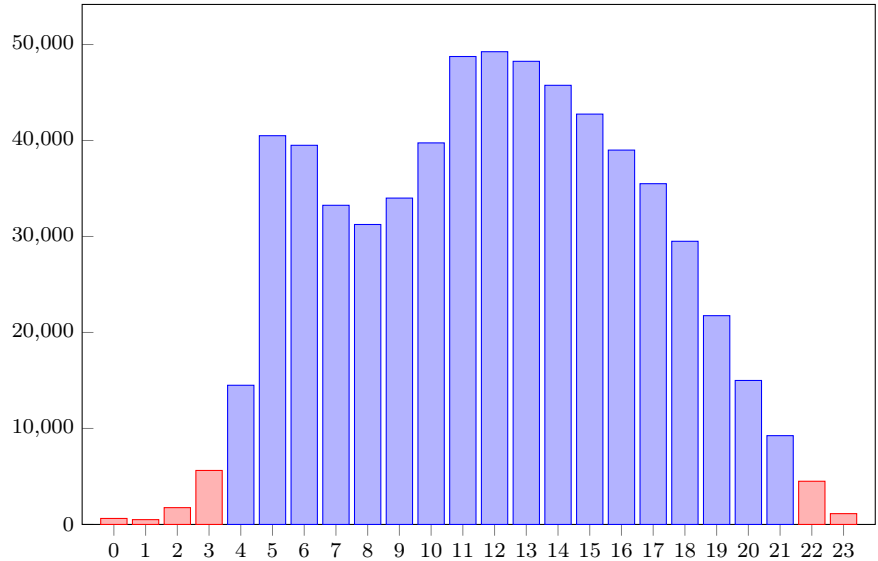
\begin{figure}[t]
	\centering\footnotesize
	\begin{tikzpicture}
		\begin{axis}[ybar, ymin = 0, xmin=0,xmax=25,width=\textwidth, height=8.5cm, bar width = 0.35cm, bar shift=0pt, scaled ticks = false, tick pos = left,
			xticklabels={0,1,2,3,4,5,6,7,8,9,10,11,12,13,14,15,16,17,18,19,20,21,22,23},xtick={1,2,3,4,5,6,7,8,9,10,11,12,13,14,15,16,17,18,19,20,21,22,23,24},
			legend style={at={(axis cs:0.5,45)},anchor=south west}]
			\addplot coordinates {
				(5,14500) 
				(6,40500) 
				(7,39500) 
				(8,33250) 
				(9,31250) 
				(10,34000) 
				(11,39750) 
				(12,48750)
				(13,49250)
				(14,48250)
				(15,45750)
				(16,42750)
				(17,39000)
				(18,35500)
				(19,29500)
				(20,21750)
				(21,15000)
				(22,9250)
			};
			\addplot coordinates {
				(1,625) 
				(2,500) 
				(3,1750) 
				(4,5625) 
				(23,4500) 
				(24,1125) 
			};
			%\legend {\emph{Bus service}, \emph{Ridepooling}};
		\end{axis}
	\end{tikzpicture}
	\caption{Hourly Distribution of W-LAN registered WSW bus rides in the city of Wuppertal during the first half of 2019.}
	\label{fig:bus_rides_per_hour}
\end{figure}

Recently, ridepooling services have emerged in many large and medium-sized cities. The idea of ridepooling is to aggregate transport requests, submitted via a smartphone app, resulting in partly shared rides. In the city of Wuppertal, Germany, \hma\footnote{\url{https://www.holmich-app.de}} was introduced in 2019 to reduce congestions and public transportation costs, as well as to improve the mobility in Wuppertal.
In the evening hours, it is necessary to offer a transport option, but buses are hardly working to capacity. Figure~\ref{fig:bus_rides_per_hour} shows the number of bus rides per hour during the first half of 2019. It is obvious that in between 10:00 PM and 3:59 AM the actual number of passengers (marked in red) is lowest, because of equally low supply and demand. For example, between 2:59 AM and 3:59 AM, there are approximately 600 requests, i.e., on average 23 requests a day during the first half of 2019. %We therefore pose the question here, whether a ridepooling service could provide a better transportation service in these evening hours. 
%\begin{figure}[t]
%	\centering
%	\includegraphics[width=\textwidth]{Visualizations/rides_hourly}
%	\caption{Hourly Distribution of W-LAN registered WSW bus rides in the city of Wuppertal \dg{during the first half of 2019}.}
%	\label{fig_rides_Wuppertal}
%\end{figure}

 This study investigates the advantages and disadvantages of (partially) substituing bus trips during periods of low demand by ridepooling services with respect to criteria related to service quality. We exemplary conduct the study in the city of Wuppertal (a mid-sized city with about 350,000 inhabitants) and substitute the bus trips in the evening hours between 10 p.m.\ and 4 a.m.\ by the \hma ridepooling service. This includes, for example, the transportation time, the waiting time, and the (excess) ride time as compared to using a private car. Since this study does neither address social criteria and constraints nor the acceptence of ridepooling services in the general public, the results have to be assessed with care.

The \hma ridepooling service operates electrical cabs with six seats and, in contrast to bus services, is not tied to fixed line plans and schedules.  \hma originated as part of a larger project, called bergisch.smart\footnote{https://www.bergischsmartmobility.de/en/the-project/}. In this context, the ridepooling service in Wuppertal was started by the local public transport provider WSW mobil, who also operates the bus service. The collaboration within the larger project enabled us to get access to ridepooling as well as bus service data. Using ride statistics recorded by logging data of WSW mobile Wi-Fi users, we model and simulate evening trip scenarios, taking into account the influence of the weekday and the full hour of a trip, as well as information on school holidays. 

Transport requests and the destination of rides are modeled statistically in a three step procedure. We first evaluate the number of requests on the basis of mobile logging data for each bus stop in the ridepooling area with destination in the same area. This leads to a Poisson count regression model for each bus stop with the hour, the weekday and an indicator for school holidays as covariates that influence the number of requests. In a second step, the number of persons traveling jointly on the trip is simulated using the statistics of \hma transport requests. Ultimately, we model the probability of a specific destination for the trip on the basis of logging out data from the mobile internet data using one multicategorical logistic regression model with the aforementioned covariates per bus stop.

On the basis of our statistical models, we simulate transport requests in the area of the \hma ridepooling service by the Monte Carlo method. We thereby generate stochastic scenarios of transport requests that approximately correspond to those of the real world transportations by bus.

%We evaluate the \dg{service quality} of shared evening rides in comparison to line based bus services by modeling the routing decisions of the ridepooling service as a dynamic dial-a-ride problem (DARP) and apply the rolling-horizon algorithm of \cite{Gaul2021}, enhanced by a feasible-path heuristic to reduce computation time, to a sufficiently large number of such simulations. 
In this simulation study we analyze the effects on the general service quality when replacing bus lines in the late evening hours in the city of Wuppertal, Germany by a ridepooling service.
The problem of the assignment of transportation requests to vehicles and the determination of vehicle routes is referred to as dial-a-ride problem (DARP).
We assume transport requests to arrive throughout the day, in contrast to being known in advance, and all information to be known with certainty, once it is received. 
This specifications make the problem a \emph{dynamic} and \emph{deterministic} DARP. In the following, we give a brief review of the related literature. %and introduce the general problem setting, including modeling assumptions and constraints.
\subsection{Related Work}
% In this simulation study we analyze the effects on the general service quality when replacing bus lines in the late evening hours in the city of Wuppertal, Germany by a ridepooling service.
% The problem of the assignment of transportation requests to vehicles and the determination of vehicle routes is referred to as dial-a-ride problem (DARP).
% We assume transport requests to arrive throughout the day, in contrast to being known in advance, and all information to be known with certainty, once it is received. 
% This specifications make the problem a \emph{dynamic} and \emph{deterministic} DARP. In the following, we give a brief review of the related literature and introduce the general problem setting, including modeling assumptions and constraints.
 
 An exhaustive survey on the DARP covering papers published since 2008 can be found in \cite{Ho2018}. For a literature review on the DARP prior to 2008 we refer to \cite{Cordeau2007}. 
 To solve the dynamic DARP in this paper we use an adaptation of the  rolling-horizon algorithm proposed in \cite{Gaul2021} which is based on the iterative solution of a state-of-the-art MILP introduced in \cite{gaul22event}.

 In the following, we focus on simulation studies concerned with the dynamic and deterministic DARP.  
 
 A review of simulation studies dealing with individual and agent-based de\-mand-responsive transport systems can be found in \cite{Ronald2015}. As the authors state, a majority of simulations are concerned with the optimization of trips, usually from an operator's perspective. In agent-based modeling, the interactions between operators and customers are studied by replicating the decision making process of individual travelers concerning the choice of destination, mode and route. 
 The first part of this review deals with the former: studies which simulate changes in features of ridepooling services from an operator's perspective, i.e., changes in parameter settings or modes of operation such as the level of dynamism. In the second part of this review, studies concerned with the impact of ridepooling on other modes of transport such as cars or public transport, and their interdependencies, are discussed. A majority of these studies is conducted using agent-based modeling.
 Two case studies are conducted in \cite{Colorni2001}: On the one hand, the authors investigate a particular problem occuring in the city of Crema in Nothern Italy: At days with a farmers' market, every customer accepted by a dial-a-ride service has to be served twice, from home to the market and back, and all trips share a common origin or destination: the market. On the other hand, they study the feasibility of a mixed static-dynamic dial-a-ride system. The authors give insights on the dependency of the level of service on the number of customers (and the number of overlapping time windows), the planning horizon and the number of vehicles and their capacity.
 In \cite{Quadrifoglio2008} the effect of a zoning vs.\ no-zoning strategy and the length of time windows in dial-a-ride services are analyzed using data from a Los Angeles ridepooling service.
 According to the study, larger time windows and a centralized dispatching system reduce the number of vehicles and miles driven but also reduce the  service quality for the %are not in favor of the 
 users of ridepooling services. 
 The effects of a partially dynamic environment are investigated in \cite{Wong2012}: The authors introduce the \emph{degree of dynamism} and investigate the influence of the ratio of dynamic requests on the system performance. Higher transportation costs and fewer accepted requests are the result of partially dynamic requests as compared to fully static or dynamic requests.
 In \cite{Haell2012} a graphical user interface is established to simulate dynamic dial-a-ride services with multiple fleets, different vehicle capacities, schedules and depots. As an illustration, the authors include costs for waiting time and users' regret in the service, and calculate the price of efficiency improvements  in exchange for less user convenience. 
 Another simulation study is conducted in \cite{Haell2015}: The authors analyze which changes in parameter settings in dynamic DARPs have a large impact on performance criteria such as customer satisfaction and operational costs and establish guidelines to service providers as a result of their study.
 In \cite{Lois2017} the trade-off between long-term highly optimized versus myopic optimization procedures is investigated and it is suggested that both optimization techniques should be used dependent on the load scenario. 
 In \cite{Hungerlaender2021} the pooling rates of an Austrian mobility provider in a rural area are aimed to be improved. In the study a large neighborhood search is implemented to solve the respective dynamic DARP and identify the most promising parameter settings to improve pooling and user convenience.
 We now summarize the literature on simulation studies which analyze the interdependencies of ridepooling with other modes of transport such as cars or public transport.
 The integration of fixed public transport and dial-a-ride services is examined in  \cite{Posada2020}. The purpose is to reduce costs of the often highly subsidized on-demand service, by allowing certain parts of the user's trips to be replaced by public transport. The authors compare the integrated with the non-integrated on-demand service and conclude that the driven distance can be reduced by 16\% using the integrated service. The proposed meta-heuristic framework can help policy makers to gain insights into the effects of an integrated service. 
 The substitution of all trips made by private cars and buses by autonomous shared vehicles in an urban setting is investigated in \cite{OECD2015}. The findings are that 9 out of 10 cars could become obsolete, resulting in a huge amount of freed space. As a negative effect, the total travel volume increases. Moreover, mixing a fleet of shared vehicles with private cars will not result in the same benefits as a pure system of autonomous shared vehicles and autonomous taxis. The simulation uses an agent-based model and synthetic trips are based on real trips generalized to a grid. 
 	The two following publications use MATSim (\cite{Axhausen2016}) as the simulation
 	software. Its basic concept is the simulation of agents that make one or more trips a day using various transport modes (e.g., car, taxi, ridepooling or public transport).
 	In \cite{Bischoff2017} the integration of shared rides into a simulation framework for non-shared taxi services is described. The simulation suggests that 15--20\% of vehicle kilometers can be saved while travel time increases at most by 3\% on average. The authors further remark that pooling works best in areas with a high taxi demand. The overall demand for pooled rides could increase even more with the introduction of autonomous vehicles since then lower fares could be offered. 
 	The complete replacement of public transport services in a mid-sized city of 100,000 inhabitants by ridepooling services is simulated in \cite{Bischoff2019}. The authors distinguish between a stop-based transportation, where the remaining distance to the customer's location is walked on foot, and door-to-door transportation. Results suggest that the current public transport system could be replaced by 300 to 400 vehicles. 
 	In \cite{Wilkes2021} the travel demand model mobiTopp is used as an agent-based simulation system. The authors describe the integration of ridesourcing, i.e., services connecting drivers of shared and non-shared taxi services with users, into the travel demand model. They analyze the impact on service providers in terms of occupation rate or number of vehicles and the interdependencies with other modes of transport. 
 	The integration of autonomous taxi services into a microsimulation is described in \cite{Dandl2017}. Travel times are modeled taking into account delays due to left turns or traffic lights. Moreover, taxi movements influence the flow in the street network and have the potential to change travel times.  The authors analyze the impact of these more realistic traffic conditions. The second focus is the impact of emtpy trips (i.e., movement of taxis without passengers on board). The simulation is conducted in the greater area of Munich, Germany.
 	In \cite{Richter2019} characteristics of transportation with autonomous vehicles are integrated into a macroscopic four-stage model (trip generation, trip distribution, modal split, and route assignment). A framework for modeling the impacts of autonomous vehicles on the network performance and capacity and on travel demand is presented. Moreover, the framework is used to evaluate the impact of autonomous vehicles on empty trips and ridepooling services. 

While the integration of ridepooling into existing simulation frameworks is considered in a variety of the above studies, and a comparison between different modes of transport takes place, the above literature review shows that the partial replacement of line-based public transport by ridepooling services during periods of low demand and its effects w.r.t.\ service quality have not been considered so far. Rather than that, a complete substitution of public transport by ridepooling or a comparison between ridepooling and taxi services is implemented. Furthermore, the generation of simulated requests calibrated to real-life bus data by means of a predictive simulation is not described in the existing literature so far.  Also, in most of the above studies, simple insertion algorithms are used to assign requests to ridepooling vehicles, which could be improved by a more sophisticated online algorithm, since the service quality of ridepooling services depends (next to the choice of service parameters such as the time window length) on the routing decisions of the underlying algorithm.

\paragraph{Contribution}
In this paper, we compare the service quality of using a bus service as compared to a dial-a-ride service during late evening hours. For this purpose we generate a statistical model of artificial transport requests, using half-a-year of log in and log out data from Wi-Fi users in buses provided by WSW. The transport requests are generated using a predictive simulation. To model the demand at each bus stop, a generalized linear model is combined with a Poisson count regression. To deal with a high density of requests, the rolling-horizon algorithm which was previously introduced by \cite{Gaul2021}, is extended by a \emph{feasible-path} heuristic to further reduce its computation time. The service quality of the dial-a-ride service is evaluated using performance measures such as transportation time, waiting time, ride time and regret (the excess ride time compared to taking a private car). This is the first study which generates transport requests using a predictive simulation and compares the service quality of buses with the service quality of dial-a-ride services in periods of low demand.

The remainder of this paper is structured as follows: %A short literature review on simulation studies in the context of the dynamic dial-a-ride problem and a description of the dynamic dial-a-ride problem is given in Section~\ref{sec:prob}.  
Section~\ref{sec:data} deals with the statistical modelling and simulation of transportation data. The optimization model used to solve instances of the dynamic dial-a-ride problem is illustrated in Section~\ref{sec:model}.  After all prerequsities have been made, a computational study evaluating the effects of the replacement of busses by ridepooling cabs with respect to the service quality is conducted in Section~\ref{sec:experiments}. Finally, based on the computational results, some conclusions are drawn in Section~\ref{sec:concl}. Here, we also comment on the shortcomings of our simulation approach and give hints for future research.

%\section{Dynamic Dial-A-Ride Routing Problem}\label{sec:prob}
%\input{prob}

\section{Statistical Modelling and Simulation of Transportation Data}\label{sec:data}
In this section we describe the simulation environment we use to generate realistic transport requests. In order to achieve this, we analyse and connect various public domain and proprietary data sets. On these data sets we base a statistical modeling of rides which so far have been conducted using the public bus system at Wuppertal. This is, among other data sources, based on anonymized Wi-Fi logging data provided by the public transportation agency WSW for the time span from January 1st to June 30th 2019. In the following, we first describe the data sources and thereafter describe the statistical modeling and simulation approach.

%We achieved a predictive simulation of new scenarios where the following important characteristics are considered: the starting station, full hour, weekday and school holidays. The simulation preserves the dependencies between these parameters and generates new scenarios which are similar to the existing ones. For each starting station we construct a Poisson model for the boardings corresponding to the selected combination of full hour, weekday and school holidays. We apply the multinomial logistical regression to guess the destinations of all trips. Moreover, we generate group trips for 1-6 passengers. The simulation results are visualized on a Leaflet-based HTML map.

\subsection{Description of the data sets}
Our statistical analysis and modeling is based on the following data sets:
\begin{itemize}
\item \emph{WSW-LAN data} consisting of information on the Wi-Fi usage by public transport passengers within the first half of the year 2019. We utilize this data to model the transport requests depending on the bus station of departure, the hour of the day, the weekday and public school holidays. This data is collected from logging information to the Wi-Fi in the WSW busses. The data is anonymized and not proprietary and has been provided by the WSW under the collaboration of the bergisch.smart.mobility consortium.
\item To provide the information, whether a day is public school holiday or not, we linked the data of \emph{School holidays}\footnote{\url{https://www.feiertagskalender.ch/export.php?geo=3069&jahr=2019&klasse=3&hl=en}} in the state of North Rhine-Westphalia with the dates in the WSW-LAN data set.
\item \emph{VRR public transport timetables}\footnote{\url{https://www.opendata-oepnv.de/ht/de/organisation/verkehrsverbuende/vrr/openvrr/datensaetze}} in General Transit Feed Specification (GTFS) format. This dataset has been utilized to analyze regret times (waiting times) by finding the predecessor bus connecting the same bus stops in the time table data. 

\begin{figure}[t]
\begin{center}
\includegraphics[width=0.85\linewidth]{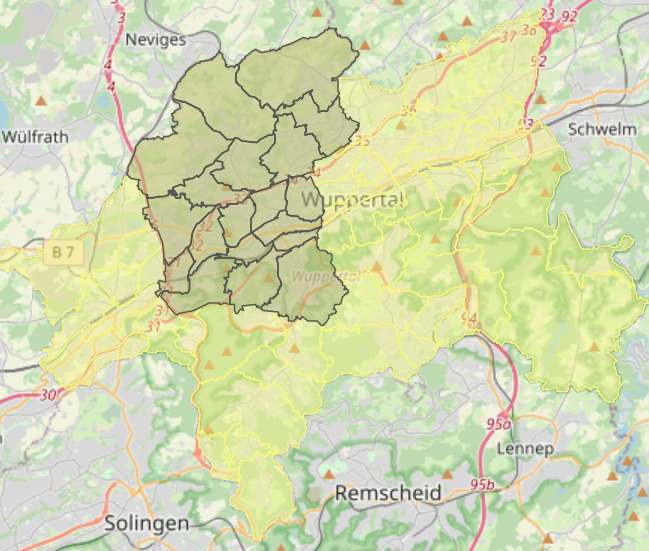}
\caption{Service area of \hma}
\label{fig_Wuppertal_and_ride_hailing_districts}
\end{center}
\end{figure}

\item  \emph{Wuppertal districts}\footnote{\url{https://www.offenedaten-wuppertal.de/dataset/quartiere-wuppertal}} as spatial data in the coordinate system EPSG:25832, see Figure~\ref{fig_Wuppertal_and_ride_hailing_districts}. This geostatistical data has been used to determine whether geocoordinates associated with bus stops are in the ridepooling operation district of the \hma. In this way, we filtered out only those bus rides that where inside the  region where \hma was available.   
\item The propriatary \hma\footnote{\url{https://www.holmich-app.de/}} data set consisting of all ridepooling requests in 2021 has been provided by \hma for the purpose of this study. We use this data to statistically model the group size of a single transport request and assimiliate it to the distribution of sizes observed in the \hma data set.
\end{itemize}

%From the data collection \href{https://www.opendata-oepnv.de/ht/de/organisation/verkehrsverbuende/vrr/openvrr/datensaetze}{\emph{VRR public transport timetables}} we have extracted the geographical coordinates of the considered stations and the waiting times between arrivals of two consequent transport vehicles at the same station.

%Using the data set \href{https://www.offenedaten-wuppertal.de/dataset/quartiere-wuppertal}{\emph{Wuppertal districts}}, we have reduced the set of all public transportation stations in Wuppertal to the service area of \href{https://www.holmich-app.de/}{\hma} (see Figure \ref{fig_Wuppertal_and_ride_hailing_districts}).

The following variables of the WSW-LAN data set are the most relevant ones for our purposes:
\begin{itemize}
\item \emph{Time of the trip start/stop}\quad (named StartTime/StopTime),
\item \emph{Station of the trip start/stop }\quad (named StartStation/StopStation).
\end{itemize}
From these variables we extracted the date, the weekday and the full hour of the trip start, and we added the school holiday information.

%\begin{figure}[H]
%\begin{center}
%\includegraphics[width=\linewidth]{Visualizations/rides_hourly}
%\caption{Simulation scenario}
%\label{fig_rides_hourly}
%\end{center}
%\end{figure}

\subsection{Statistical modeling of transport requests}

Generalized linear models (GLM) are the standard models of statistical analysis beyond standard regression, see e.g.\ \cite{Faraway2016,hastie2009elements,knight1999mathematical}. As transportation events at bus stops, i.e., a person taking a bus at a specific time at a given hour, are counted by integer values, the standard regression based on the continuous normal distribution is inadequate. Here, we therefore choose a GLM-based \emph{poisson count regression} approach, which we now shortly explain: 

Let $x_i$ be the covariate (daytime, hour and an indicator of school holidays in our case) and $y_i$ the actual number of rides observed at a particular bus stop using the Wi-Fi logging information and geolocation of the bus that is matched to geolocations of the bus stops. A common assumption in count regression is to model the corresponding random number $Y_i$ from which $y_i$ is a sample, as Poisson distributed 
\[
P(Y_i=y)=\textup{Po}(\lambda_i)(\{y\})=e^{-\lambda_i}\,\frac{\lambda_i^y}{y_i!}~,\qquad y=0,1,2\ldots,
\]       
where $\lambda_i$ is the intensity parameter or the expected value of requests for the instance $i$. In GLM this intensity parameter is modeled using a link function that maps the expected value to a link variable which is modeled as a linear combination of the covariates plus an intercept. For Poisson count regression, it is common practice to use the canonical link function $\log(\cdot)$, hence
\[
\log(\lambda_i)=\beta^\top x_i \iff  \lambda_i=\exp(\beta^\top x_i).
\]

Inserting this to the Poisson distribution enables one to determine the parameters $\beta\in\mathbb{R}^q$ via Fisher scoring maximizing the likelihood, see \cite{knight1999mathematical}. Note that $x_i$ is composed of one intercept term, $6$ dummy $0$-$1$ variables indicating the weekdays Tuesday--Sunday (with all those dummies equal zero indicating Monday) and $23$ dummy variables for the hours of daytime different from 12--1\,PM. The set of covariates is complemented with a further dummy variable for school holidays, so that we obtain $q=1+6+23+1=30$ parameters that are estimated from the data for each of the 516 bus stops in the \hma ridepooling area seperately. This is automated by adequately filtering and accumulating the requests and application of the \texttt{glm} function in the statistics programming language \texttt{R}\footnote{\url{https://r-project.org}}.  

In this way we estimate the distribution of raw transport requests on the basis of single passengers. While this adequately models single bus passengers, it does not model the group size of a ridepooling request. To obtain a request model with adequate group sizes, we use the statistics of group sizes of the ridepooling requests of the \hma as obtained from the corresponding data set and displayed in Table \ref{tab:group_sizes}.
%To model the boardings at each station during any full hour, we naturally assume that boardings in disjoint time intervals are independent. Our approach considers group rides, too, and we assume that the different groups have different boarding times (practically, the hour, the minutes, the seconds and the milliseconds of the boarding times of two different groups cannot be all identical). Moreover, we assume that the boarding rate within each full hour does not change essentially. The latter assumption could seem even more reasonable for shorter time intervals, but the disadvantage here would be the smaller set of observations, not sufficient for stochastic simulations.

%The above assumptions imply that the number of boardings can be modelled as a \emph{Poisson counting process} (see, e.g., \cite[Theorem 8.1]{Goodman2006}).

%Based on the preselected values of weekday, full hour and school holiday information, we first predict the boarding intensities of each outbound station, using the Poisson model with the log-link. According to WSW mobile statistics, the number of actual passengers is about $3.5$ times larger than that of WSW-LAN users. Therefore we upscale the intensities by the factor $3.5$.\todo[inline]{DG: sollen wir das raus nehmen und das Paper so lassen (1-fache Simulation)?} In the next step we generate the group sizes whereby considering group rides consisting of maximum 6 passengers. The sample probabilities for group sizes are extracted from the \hma data. E.g., the group size distribution for the current run looks like this:

\begin{table}[H]
\begin{centering}
\begin{tabular}{ccccccc}
\toprule
Group size & 1 & 2 & 3 & 4 & 5 & 6\\
\midrule
Probability & 0.804 & 0.153 & 0.026 & 0.011 & 0.004 & 0.002\\
\bottomrule
\end{tabular}
\par\end{centering}
\caption{Distribution of passenger group sizes}
\label{tab:group_sizes}
\end{table}

In order not to increase the average overall number of passengers, downscale the  intensities $\lambda_i$ by the average passenger group size (i.e., by $1.264$ in the case of Table \ref{tab:group_sizes}).  After using the downscaled intensities for generation of a statistically realistic random number of requests, for each request the size of groups are generated according to their distribution.

\subsection{Statistical modeling of destinations}

To simulate the destinations, we use the multinomial logit model, see \cite[Sec. 7.1]{Faraway2016}. Based on the same set of $30$ dummy covariates $x_i$ as described in the previous subsection, we train a $(s-1)\times q$   $\Theta_i$ matrix (with $q-1$ the number of covariates), which generates an   $s-1$ dimensional activation vector $z_i=z(x_i|x_i,\Theta_i)$. Thereby, $s$ denotes the number of bus stations in the ridepooling operation area that have ever been reached from the starting bus station under consideration.  One frequent destination bus station from the given bus station of departure is selected, put at the first index position in $z_i$ and associated with intensity $z_{i,1}=0$ to avoid overparametrization. We thereafter send the entire activation scores $z_i$ through the softmax activation function
\[
p(j|x_i,i,\Theta_i)=\textup{softmax}(z_i)_k=\frac{\exp(z(x_i,\Theta_i)_j)}{\sum_{l=1}^s\exp(z(x_i,\Theta_i)_l})~~\text{for} ~~j=1,\ldots, s,
\]    
which gives us the vector of discrete probabilities of choosing a specific destination among the $s$ existing options.

% For the preselected simulation characteristics, i.e., for the weekday, for the full hour of the trip start and for the school holiday information we read all possible destinations with their probabilities from the original WSW-LAN data set. Let $\mathcal{H}$ denote the set of all stations. For two different
%stations $A$ and $B$, the probability $p(A,B)$ of traveling from
%station $A$ to station $B$ is modeled to be equal to
%\[
%\frac{\exp\beta(A,B)}{\sum\limits _{C\in\mathcal{H}\backslash\{A\}}\exp\beta(A,C)}
%\]
%with suitable scaling factors $\beta(A,C)$. Assuming the dependence
%of quantity $\beta(A,C)$ on $C$ to be linear, we predict its value
%by fitting neural networks with the aid of the function \emph{multinom}

For each of the 516 stations of departure, the above multicategorical logistic regression (or equivalently a shallow neural net without hidden layer) is fitted using  the \texttt{R} package \texttt{nnet} after an adequate filtering for the allowed destinations.
Note that due to the shortcomings of anonymous Wi-Fi tracking data, no bus rides requiring a transfer can be modeled with our approach. 

\subsection{Simulation of virtual data sets}
Based on the statistical models for transport requests, group sizes and destinations, we are now able to sample from the corresponding distributions in order to create a virtual request scenario. In detail, this is done for every station simulating an entire week, hour by hour, either for school holidays or not. Within the given hours, the exact time of the transport requests is uniformly distributed.    

\begin{figure}[t]
\includegraphics[width=\linewidth]{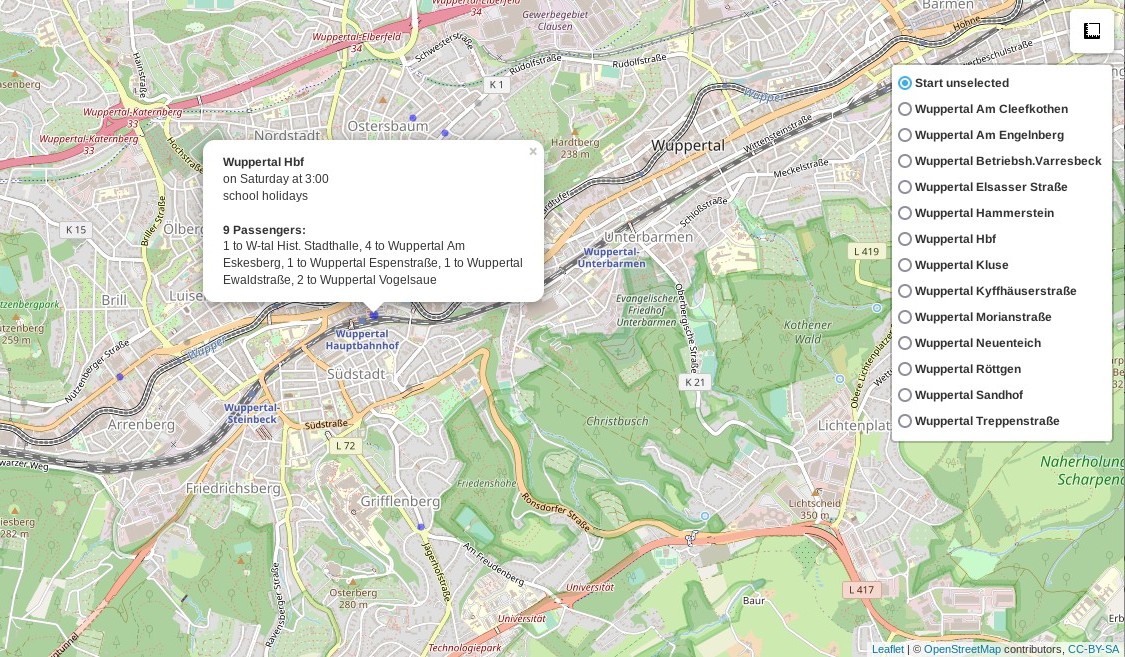}
\caption{Visualization of a ridepooling scenario on the interactive map.}
\label{fig:simulationscenario}
\end{figure}

We also provide an interactive tool for visualization of the request scenarios.The user can select the outbound time (in full hours), the day of the week, and the school holiday information. This results in a Leaflet-based HTML map where stations are displayed as disc-shaped markers with radii proportional to the square root of the corresponding entry number. Hovering over a marker displays the name of the station, and clicking on it displays boarding information, i.e., number of boardings, start time (in full hours), day of the week, school holiday information, and destination stops. For a better view, one can select the outgoing station in the right panel, then the corresponding destinations will be displayed by connecting rays. We use blue color for all the stations where there are boarding passengers, whereas the stations where no passengers get on a bus are displayed using the grey color, see Figure 
\ref{fig:simulationscenario} for an example scenario.

\subsection{Modeling of waiting times}
\label{sec:waitingtimes}
To model waiting times for bus rides, we analyze the \emph{VRR public transport timetables} by searching a preceding trip in the time tables that directly connects the bus stops of origin with the destination which gives us the maximum waiting time. If the maximum waiting time exceeds a threshold of two hours, the value is set to the threshold. The waiting time is now simulated uniformly in between the maximum waiting time and zero, as, potentially, the passenger might have wanted to travel at any of these times and we have no reason to assume a preference for any point of time in the given time interval.

\section{Optimization Model}\label{sec:model}
\noindent

To solve an instance of the dynamic DARP we use the rolling-horizon algorithm by \cite{Gaul2021} enhanced by a \emph{feasible-path heuristic.}  As opposed to most heuristic approaches to the dynamic DARP (see \cite{Gaul2021}, \cite{Ho2018}), the rolling-horizon algorithm consists of the iterative computation of exact optimal solutions subject to given prior routing decisions. To further reduce computation times in order to compensate a high density of requests, the algorithm is extended by a feasible-path heuristic. By means of the heuristic, the complexity of the MILPs that are solved iteratively can be controlled. Thus, the solution strategy we use is easily adjustable to different intensities of requests, giving way to obtain near-optimal solutions when request density is low, and speeding up computation times when it is high. After a description of the most important parameters used, a short summary of the algorithm is given below, followed by a description of the newly added feasible-path heuristic. 

In the following mathematical formulation of the dynamic DARP, we use the terminology of \cite{Gaul2021}. A series of $n$ transport requests is either answered and integrated into vehicle routes, or denied. An answer has to be communicated to a request within $\Delta$ seconds. We assume the requests to be ordered increasingly according to the time $\tau_i - \Delta$ they are revealed, i.e.\ $\tau_1 - \Delta \leq \ldots \leq \tau_n - \Delta$. A transport request $i\in\{1,\ldots,n\}$ consists of a pick-up location $i^+$, a drop-off location $i^-$, a desired earliest pick-up time $e_{i^+}$ and  a fixed number of passengers to be transported. We assume a service time of $s_i \geq 0$ minutes (which is required to enter or leave the vehicle). There is a vehicle fleet of fixed size and capacity $Q$ situated at the vehicle depot $0$ ready to serve the requests. The travel time and routing cost between two locations $j_1, j_2 \in \{0, 1^+, \ldots, n^+, 1^-, \ldots, n^-\}$ is denoted as $t_{j_1j_2}$ and $c_{j_1j_2}$, respectively.  From the earliest desired pick-up time $e_{i^+}$, a pick-up time window $\left[e_{i^+}, \ell_{i^+}\right]$ and a drop-off time window $\left[e_{i^-}, \ell_{i^-}\right]$ are computed, taking into account service and travel times. The time window $\left[e_0, \ell_0\right]$ at the depot describes the overall start $e_0$ and end $\ell_0$ of service; the vehicles may not leave the depot before $e_0$ and may not arrive back at the depot later than $\ell_0$. If a request is assigned to a vehicle route, its pick-up and drop-off location have to be contained in the vehicle route in the correct order, and pick-up and drop-off have to take place within the time windows.  The vehicle capacity may not be exceeded at any time. The time needed to transport request $i$ may not exceed a prespecified ride time.  Upon accepting a request, a pick-up time is communicated, and may not be postponed by more than a fixed time.  A list of all parameters used in this paper can be found in Table~\ref{paras}.

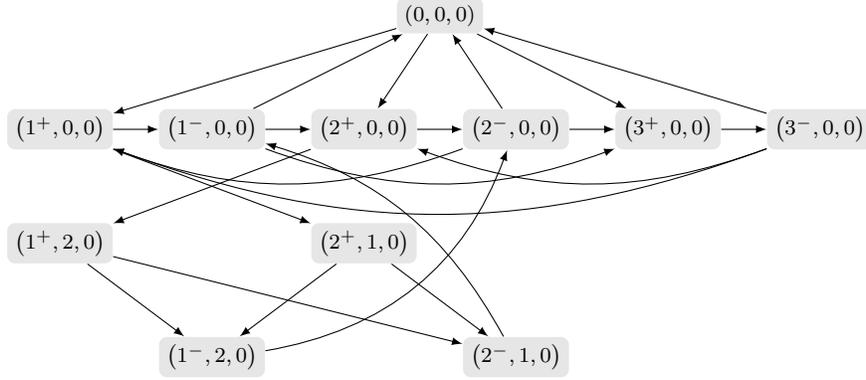
\begin{figure}
	\begin{scaletikzpicturetowidth}{\textwidth}
		\footnotesize
		\begin{tikzpicture}[scale = \tikzscale, every node/.style={rectangle,fill=black!10!white},rounded corners=3]
			%[scale = \tikzscale,]
			%[xscale=0.6,yscale=0.58,every node/.style={draw,ellipse}]
			\useasboundingbox (-12,-11) rectangle (12,0.75);
			% Ovale um nodes!
			\node (null) at (0,0) {$(0,0,0)$};
			\node  (100) at (-10,-3) {$\left(1^+,0,0\right)$};
			\node (200) at (-2,-3) {$\left(2^+,0,0\right)$};
			\node (300) at (6,-3) {$\left(3^+,0,0\right)$};
			\path (null) edge[-latex] (100);
			\path (null) edge[-latex] (200);
			\path (null) edge[-latex] (300);
			\node (400) at  (-6,-3) {$\left(1^-,0,0\right)$};
			\node (500) at (2,-3) {$\left(2^-,0,0\right)$};
			\node (600) at (10,-3) {$\left(3^-,0,0\right)$};
			
			\path
			(100) edge[-latex] (400)
			(200) edge[-latex] (500)
			(300) edge[-latex] (600);
			
			\path
			(400) edge[-latex] (null)
			(500) edge[-latex] (null)
			(600) edge[-latex] (null);
			
			\path
			(400) edge[-latex] (200)
			(400) edge[-latex, bend angle = 20, bend right] (300);
			
			\path
			(500) edge[-latex, bend angle = 20, bend left] (100)
			(500) edge[-latex] (300);
			
			\path
			(600) edge[-latex, bend angle = 20, bend left] (100)
			(600) edge[-latex, bend angle = 20, bend left] (200);	
			
			%\node (210) at (-8,-6) {$\left(2^+,1,0\right)$};
			\node (210) at (-2,-6) {$\left(2^+,1,0\right)$};
			
			\path
			(100) edge[-latex] (210);
			
			%\node (120) at (0,-6) {$\left(1^+,2,0\right)$};
			\node (120) at (-10,-6) {$\left(1^+,2,0\right)$};
			
			\path
			(200) edge[-latex] (120);
			
			\node (420) at  (-6,-9) {$\left(1^-,2,0\right)$};
			%\node (510) at  (-2,-9) {$\left(2^-,1,0\right)$};
			\node (510) at  (2,-9) {$\left(2^-,1,0\right)$};
			
			\path 
			(210) edge[-latex] (420)
			(210) edge[-latex] (510);
			
			\path 
			(120) edge[-latex] (420)
			(120) edge[-latex] (510);

			\draw (510) edge[-latex, bend angle = 22, bend right] (400);
			%\draw (420) edge[-latex, bend angle = 10, bend left] (500);
			\draw (420) edge[-latex, bend angle = 30, bend right] (500);
			
		\end{tikzpicture}
	\end{scaletikzpicturetowidth}
	\caption{Graph representation of an example with three users\label{fig:example} taken from \cite{gaul22event}.}
\end{figure}

In the rolling-horizon algorithm  a series of MILP formulations MILP($\tau_\ell$),  $\ell=1,\ldots,n$ is solved.  Each formulation corresponds to a subproblem DARP($\tau_\ell$) of the dynamic DARP, in which only the requests that have arrived up to time $\tau_\ell - \Delta$ are known. Requests that have been denied or dropped-off up to time $\tau_\ell$ do not have to be considered in DARP($\tau_\ell$). In addition, MILP($\tau_\ell$) incorporates the routing decisions that have been made up to $\tau_\ell$ based on the solutions computed in prior iterations. The underlying structure of the MILP formulations is the dynamic event-based graph  $G(\tau_\ell)$ from \cite{Gaul2021}.
The node set of $G(\tau_\ell)$ consists of $Q$-tuples that represent events which correspond to feasible user allocations in the vehicle, given the current time $\tau_\ell$. More precisely, each component of a node $v$ contains a request index representing a user with all related data, or the value $0$. The first coordinate of a node $v$ contains additional information about the location, i.e., the depot or a pick-up or a drop-off location of the associated pick-up or drop-off event corresponding to  time $\tau_\ell$. The arc set consists of feasible transitions between these events. An illustration can be seen in Figure~\ref{fig:example}, and a detailed description of the dynamic event-based graph can be found in \cite{Gaul2021}.

Only nodes which represent allocations of users that are pairwise feasible in the following sense are allowed in the graph: 
A user $i$ is pairwise feasible with user $j$, if the paths $j^+\rightarrow i^+ \rightarrow j^- \rightarrow i^-$ or $j^+ \rightarrow i^+ \rightarrow i^- \rightarrow j^-$ are feasible w.r.t.\ time window and ride time constraints. Every time a new user $i$ arrives, the pairwise feasibility of $i$ with all other current users $j$, i.e.\ all users which have arrived up to time $\tau_\ell-\Delta$ and have not been denied or dropped-off yet, is checked and the graph is modified accordingly. Moreover, only nodes which represent allocations of users in the vehicle which are feasible w.r.t.\ vehicle capacity are allowed. 
To take into account service providers' as well as users' interests the objective function in MILP$(\tau_j)$ is a weighted-sum objective with three criteria:
 \begin{equation*}
 	\omega_1 \cdot \text{ routing costs} + \omega_2 \cdot\text{ regret} + \omega_3\cdot \text{ number of denied requests},
 \end{equation*}
 where $\omega_1, \omega_2, \omega_3>0$ are positive weigths, and regret refers to the amount of time a user has to spend as compared to using a private car, including the waiting time for the ridepooling vehicle to pick up the user.
 
To account for a high density of requests, the rolling-horizon algorithm is extended by the use of a heuristic procedure: 
The feasible-path heuristic ensures that every time a new request $i$ arrives, only nodes and arcs corresponding to the $\rho$ most temporally and spatially proximate paths $j^+\rightarrow i^+ \rightarrow j^- \rightarrow i^-$ and $j^+ \rightarrow i^+ \rightarrow i^- \rightarrow j^-$ are added to the dynamic event-based graph. Only paths of pairwise feasible requests $i,j$ are considered. The spatial proximity of the path $j^+\rightarrow i^+ \rightarrow j^- \rightarrow i^-$ is measured in terms of the function
\begin{equation*}
	\omega_1 \left(c_{j^+i^+} + c_{i^+j^-} + c_{j^-i^-} \right), 
\end{equation*} 
whereas the temporal proximity is calculated as 
\begin{equation*}
	\omega_2 \left( 2 \left( \max ( e_{i^+}, e_{j^+} + s_{j}+ t_{i^+j^+} ) + s_{i} + t_{i^+j^-}\right) - e_{j^-} + s_{j} + t_{j^-i^-}- e_{i^-} \right).
\end{equation*} 
The temporal and spatial proximity of the path $j^+ \rightarrow i^+ \rightarrow i^- \rightarrow j^-$ is calculated accordingly. The temporal and spatial proximity of a path which is infeasible w.r.t.\ time window or ride time constraints is set to infinity. 

Let $\rho_i$ denote the number of feasible paths associated with a new request $i$. The number of feasible paths that are considered as a basis for the computation of new nodes and arcs in the dynamic event-based graph is denoted by $\rho = \max(\rho_{\text{abs}}, \rho_{\text{rel}} \rho_i )$, which is composed of a fixed bound $\rho_{\text{abs}}$ on the number of considered feasible paths and a relative bound $\rho_{\text{rel}}$ on the percentage of considered feasible paths.  By the use of the heuristic, the size of the graph and the number of modifications of constraints and new variables in the dynamic event-based MILP at every iteration is kept small. Nevertheless, we lose an important characteristic of the rolling-horizon algorithm: if MILP$(\tau_j)$ is solved to optimality within the time limit, it is not guaranteed that the solution returned by the MILP solver is the global optimal solution w.r.t.\ the current vehicle routes computed in prior iterations. In preliminary tests, we observed that using the heuristic with $\rho_{\text{abs}} = 10$ and $\rho_{\text{rel}} = 0.25$, average routing costs increased by 2.8\%  while average regret decreased by 0.6\% and computational time decreased by 65\%. Thus, the rolling-horizon-algorithm paired with the feasible-path-heuristic is able to produce high-quality solutions in a significantly reduced amount of time.

\section{Numerical Experiments}\label{sec:experiments}
In the first part of this section, 30 weeks of simulated ridepooling scenarios with service hours from 22:00 to 03:59 are considered and optimized tours are computed using the rolling-horizon algorithm \cite[proposed in][]{Gaul2021},  enhanced by the feasible-path heuristic presented in Section~\ref{sec:model}. To evaluate the quality of the transportation via ridepooling as compared to the bus service, we rely on the quality measures which are regularly used as parts of the objective function in ridepooling applications, see for example \cite{Ho2018} and the references given therein.  Secondly, to measure the quality of the bus service, the WSW-LAN data set is extended by simulated waiting times as described in Subsection~\ref{sec:waitingtimes} at the bus stop, and the two modes of transportation are compared.

In total, 210 simulated instances are solved, 30 instances for each day of the week. 
We use the following parameter settings for the feasible-path-heuristic and the rolling-horizon algorithm for the tour computations:
$\rho_{\text{abs}} = 10$, $\rho_{\text{rel}} = 0.25$, and $\Delta = 45$.
All other parameters are given by the conditions of transport of the service provider, i.\,e., the service time is considered to be constant and equals $s_i = 45$ seconds and the length of the pick-up time window is given by $\ell_{i^+} - e_{i^+} = 25$ minutes. The number of transport requests, the number of requested seats, and pick-up and drop-off locations are  taken from the simulated transport data. The distances between the latter are computed with help of the Python packages \texttt{OSMnx} and \texttt{NetworkX}, and the travel time between two locations $j_1,j_2$ is computed based on a linear regression using distances and travel times from the \hma data set: $t_{j_1j_2}=2.3634\cdot c_{j_1j_2} + 0.2086$. 
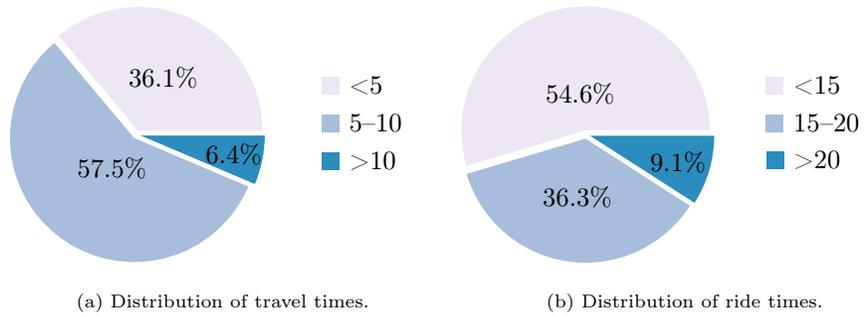
\begin{figure}
	\begin{subfigure}{0.5\textwidth}
		\begin{tikzpicture}[scale=0.55]
			\tikzset{
				lines/.style={draw=none},
			}
			\pie[text=legend, color = {bluepie1, bluepie2, bluepie3},style={lines}, explode=0.1]{36.1/ <5, 57.5/5--10,6.4/>10}
		\end{tikzpicture}
		\caption{Distribution of travel times.}
	\end{subfigure}
	\begin{subfigure}{0.5\textwidth}
		\begin{tikzpicture}[scale=0.55]
			\tikzset{
				lines/.style={draw=none},
			}
			\pie[text=legend, color={bluepie1, bluepie2, bluepie3}, style={lines}, explode=0.1]{54.6/ <15, 36.3/15--20,9.1/>20}
		\end{tikzpicture}
		\caption{Distribution of ride times.}
	\end{subfigure}
	\caption{Distribution of travel and ride times (in minutes) among the artificial requests.	\label{fig:pie}}
\end{figure}
A visualization of the distribution of travel and ride times among the the transport requests is given in Figure~\ref{fig:pie}. It remains to determine the size of the vehicle fleet.

For the purpose of this simulation we use a fixed number of vehicles depending on the weekday; the size of the vehicle fleet should not vary from week to week or throughout the evening. Hence, for each weekday, using our estimated transport requests, we calculate the total number of requests per hour over 30 weeks and take the average among the 30 weeks. Then, the hour with the maximum average number of requests is used as a basis to estimate the number of vehicles needed. Preliminary tests have shown that a ratio of 8 requests per vehicle and hour on average is a good starting point. In the following we will refer to this parameter setting as scenario A. Additionally, we simulated also with a fleet size reduced by one (scenario B) and increased by one (scenario C).
The corresponding sizes of the vehicle fleet are given in Table~\ref{tab:numveh}.

\begin{table}
	\centering
	\begin{threeparttable}
		\small
		\begin{tabular}{lcccc}
			\toprule
			& \parbox{5em}{max.\ avg.\\ \# requests per hour}  &  A & B & C \\
			\midrule
			Monday        &  37.2 & 5 & 4 & 6 \\
			Tuesday        &  34.6 & 5  & 4 & 6 \\
			Wednesday & 26.7 & 4  & 3  & 5 \\
			Thursday     & 30.8  & 4& 3 & 5 \\
			Friday           & 24.1 & 4 & 3 & 5 \\
			Saturday      & 17.9 & 3 & 2 & 4 \\
			Sunday         & 26.5 & 4 & 3 & 5 \\
			\bottomrule
		\end{tabular}		
		\caption{Number of vehicles depending on the weekday and scenario.}
		\label{tab:numveh}
	\end{threeparttable}
\end{table}

The code of the rolling-horizon algorithm is written in C++ and the MILPs are solved using IBM ILOG CPLEX~12.10. The computations are performed on an Intel Core~i7-8700 CPU, 3.20~GHz, 32~GB memory.

To analyze the service quality of the bus service in comparison to the ridepooling service, the results are evaluated using the measures for service quality listed below. For this purpose, let $B_i$ denote the beginning of service, and $D_i$ the departure time at the origin of request $i$, and $A_i$ the arrival time at the destination of request $i$. Let $p^\star\in \{0,1\}^n$ be the part of the solution vector of the last MILP solved,  MILP($\tau_n)$, where 
\begin{equation*}
	p^\star_i = \begin{cases}
		1 & \text{request $i$ is accepted,} \\
		0 & \text{else}.
	\end{cases}
\end{equation*} 
Then, the quality measures used to evaluate the service quality in this simulation are defined as follows:
\begin{itemize}
	\item average regret, i.\,e.\ \quad $\displaystyle\frac{1}{\sum_{i=1}^np^\star_i}\sum_{i=1}^{n} p^\star_i  (A_i - e_{i^-})$,
	\item average waiting time, i.\,e.\ \quad  $\displaystyle\frac{1}{\sum_{i=1}^np^\star_i} \sum_{i=1}^n p^\star_i ( B_i - e_{i^+})$, 
	\item average ride time, i.\,e.\ \quad  $\displaystyle\frac{1}{\sum_{i=1}^np^\star_i} \sum_{i=1}^n p^\star_i (A_i - D_i)$.
	\item average transportation time, i.\,e.\ \quad  $\displaystyle\frac{1}{\sum_{i=1}^np^\star_i} \sum_{i=1}^n p^\star_i (A_i - e_{i^+})$.
\end{itemize}

Additionally, we report the following quantities for the simulated bus requests transported via ridepooling, as they are criteria of our weighted-sum objective function:
\begin{itemize}
	\item total routing costs, i.\,e., total vehicle kilometers,
	\item percentage denied, i.\,e., the percentage of denied requests.
\end{itemize}
Table~\ref{tab:30weeks} illustrates the results of the simulation compared against the bus service, where for each scenario and each weekday average values over 30 weeks are reported. Additionally, in Table~\ref{tab:30mondays} we exemplarily list average hourly results for Monday evening and each scenario. In both tables, we report average regret, waiting time, ride time and transportation time for the bus service based on trip data from the WSW-LAN data set. %This data set contains trip data deduced from the Wifi-usage by passengers during the first half of 2019. 
In order to compare the quality of both transportation modes, the trip data of each passenger in the WSW-LAN data set is extended by a simulated time the passenger has to wait at the stop for the bus to arrive, see Section~\ref{sec:waitingtimes}. As for the ridepooling service, we assume a service time of 0.75 minutes for passengers to enter or leave a bus. Due to the reduced amount of passengers during the late evening hours this is a realistic assumption. Hence, for all bus passengers $i$ we have $D_i = B_i + 0.75$.  The number of buses simultaneously in service was estimated by WSW to be a minimum of 10 buses during the week. Hence we used this number for comparison in the following.

  \setlength{\tabcolsep}{7pt}
 \begin{table}[h!tb]
 	\begin{threeparttable}
 		\footnotesize
 		\begin{tabular}{rSSSSSSS[table-format=3.2]}		
 			{\tabrotate{Day}} & {\negthickspace\tabrotate{\# vehicles}} & {\tabrotate{Total routing costs}} & {\tabrotate{\% denied}} & {\tabrotate{Avg.\ regret}} & {\tabrotate{Avg.\ waiting time}}  & {\tabrotate{Avg.\ ride time}} & {\tabrotate{Avg.\ transp.\ time}} \\ 
 			\midrule
 			\textbf{Scenario A} &  &  &  &  &  &  &  \\
 			Mon & 5 & 251.7 & 0.8 & 9.6 & 8.1 & 6.9 & 15.8 \\
 			Tue & 5 & 268.9 & 0.0 & 9.5 & 8.0 & 7.0 & 15.7 \\
 			Wed & 4 & 238.5 & 0.9 & 10.3 & 8.8 & 6.8 & 16.4 \\
 			Thu & 4 &231.1 & 1.2 & 10.3 & 8.8 & 6.9 & 16.5 \\
 			Fri & 4 &217.2 & 0.3 & 9.6 & 8.2 & 6.7 & 15.7 \\
 			Sa & 3 &148.7 & 0.4 & 9.2 & 8.0 & 6.3 & 15.1 \\
 			Su & 4 & 176.8 & 0.6 & 9.7 & 8.3 & 6.8 & 15.9 \\
 			\midrule
 			\textbf{Scenario B} &  &  &  &  &  &  &  \\
 			Mon & 4 & 243.5 & 3.3 & 11.2 & 9.5 & 7.2 & 17.4 \\
 			Tue & 4 & 262.2 & 2.1 & 11.8 & 10.0 & 7.2 & 18.0 \\
 			Wed & 3 & 222.9 & 6.6 & 13.4 & 11.5 & 7.2 & 19.4 \\
 			Thu & 3 & 218.1 & 6.8 & 12.6 & 10.7 & 7.3 & 18.8 \\
 			Fri & 3  & 207.8 & 3.9 & 12.2 & 10.4 & 7.0 & 18.3 \\
 			Sa & 2 & 140.9 & 6.9 & 12.5 & 10.8 & 6.8 & 18.4 \\
 			Su & 3  & 167.7 & 5.1 & 11.6 & 9.9 & 7.1 & 17.8 \\
 			\midrule
 			\textbf{Scenario C} &  &  &  &  &  &  &  \\
 			Mon & 6 & 251.8 & 0.1 & 8.5 & 7.1 & 6.8 & 14.7 \\
 			Tue & 6 & 267.6 & 0.0 & 8.4 & 7.0 & 6.8 & 14.6 \\
 			Wed & 5 & 237.1 & 0.0 & 8.4 & 7.2 & 6.5 & 14.5 \\
 			Thu & 5  & 234.2 & 0.0 & 8.9 & 7.5 & 6.8 & 15.1 \\
 			Fri & 5  & 216.5 & 0.0 & 8.1 & 6.9 & 6.5 & 14.1 \\
 			Sa & 4 & 148.6 & 0.0 & 7.7 & 6.7 & 6.1 & 13.6 \\
 			Su & 5 & 178.3 & 0.0 & 8.1 & 7.0 & 6.6 & 14.4 \\
 			\midrule
 			\textbf{Bus service} &  &  &  &  &  &  &  \\
 			Mon       &10&&&24.1&17.7&11.2&29.7 \\
 			Tue       &10&&&24.7&19.0&10.7&30.4 \\
 			Wed       &10&&&25.1&19.0&11.0&30.7 \\
 			Thu       &10&&&22.6&16.1&11.3&28.1 \\
 			Fri       &10&&&22.4&16.8&10.5&28.0 \\
 			Sa        &10&&&19.5&13.7&10.5&25.0 \\
 			Su        &10&&&20.1&14.6&10.2&25.6 \\
 			\bottomrule 
 		\end{tabular}
 		\caption{Average computational results over 30 weeks of simulated request data and average transportation data computed over bus trips during first half of 2019.}
 		\label{tab:30weeks}
 	\end{threeparttable}
 \end{table}

 \setlength{\tabcolsep}{7pt}
	\begin{table}[h!tb]
	\begin{threeparttable}
		\footnotesize
		\begin{tabular}{rSSSSSSS[table-format=3.2]}		
			{\tabrotate{Day}} & {\negthickspace\tabrotate{\# vehicles}} & {\tabrotate{Total routing costs}} & {\tabrotate{\% denied}} & {\tabrotate{Avg.\ regret}} & {\tabrotate{Avg.\ waiting time}}  & {\tabrotate{Avg.\ ride time}} & {\tabrotate{Avg.\ transp.\ time}} \\ 
			\midrule
			\textbf{Scenario A} &  &  &  &  &  &  &  \\
			Mon & 5 & 251.7 & 0.8 & 9.6 & 8.1 & 6.9 & 15.8 \\
			\cdashlinelr{2-8}
			Mon 22-23 & 5 &  & 0.0 & 8.2 & 6.7 & 6.3 & 13.8  \\
			Mon 23-00 & 5 &  & 0.0 & 6.7 & 6.3 & 5.6 & 12.6  \\
			Mon 00-01 & 5 &  & 0.0 & 6.0 & 5.6 & 5.5 & 11.9  \\
			Mon 01-02 & 5 &  & 0.0 & 5.7 & 5.3 & 7.0 & 13.0  \\
			Mon 02-03 & 5 &  & 0.0 & 8.1 & 7.1 & 7.5 & 15.4  \\
			Mon 03-04 & 5 &  & 1.9 & 12.1 & 10.0 & 7.4 & 18.2  \\
			\midrule
			\textbf{Scenario B} &  &  &  &  &  &  &  \\
			Mon & 4 & 243.5 & 3.3 & 11.2 & 9.5 & 7.2 & 17.4 \\
			\cdashlinelr{2-8}
			Mon 22-23 & 4 &  & 0.0 & 9.3 & 7.6 & 6.4 & 14.8\\
			Mon 23-00 & 4 &  & 0.0 & 7.2 & 6.7 & 5.6 & 13.1\\
			Mon 00-01 & 4 &  & 0.0 & 6.0 & 5.5 & 5.6 & 11.9\\
			Mon 01-02 & 4 &  & 0.0 & 6.2 & 5.8 & 6.8 & 13.4\\
			Mon 02-03 & 4 &  & 0.0 & 8.8 & 7.5 & 7.8 & 16.1\\
			Mon 03-04 & 4 &  & 7.8 & 14.0 & 11.6 & 7.3 & 19.7\\
		\midrule
			\textbf{Scenario C} &  &  &  &  &  &  &  \\
			Mon & 6 & 251.8 & 0.1 & 8.5 & 7.1 & 6.8 & 14.7 \\
			\cdashlinelr{2-8}
			Mon 22-23 & 6 &  & 0.0 & 7.9 & 6.4 & 6.2 & 13.5\\
			Mon 23-00 & 6 &  & 0.0 & 6.4 & 5.9 & 5.7 & 12.3\\
			Mon 00-01 & 6 &  & 0.0 & 5.6 & 5.3 & 5.5 & 11.5\\
			Mon 01-02 & 6 &  & 0.0 & 5.4 & 5.0 & 6.7 & 12.5\\
			Mon 02-03 & 6 &  & 0.0 & 7.5 & 6.6 & 7.4 & 14.7\\
			Mon 03-04 & 6 &  & 0.4 & 10.1 & 8.4 & 7.2 & 16.4\\
		\midrule
		\textbf{Bus service} &  &  &  &  &  &  &  \\
			Mon       &10&&&24.1&17.7&11.2&29.7 \\
			\cdashlinelr{2-8}
			Mon 22-23 &10&&&24.1&17.7&11.2&29.6 \\
			Mon 23-00 &10&&&24.5&18.1&11.2&30.1 \\
			Mon 00-01 &10&&&20.1&15.0&10.1&25.9 \\
			Mon 01-02 &10&&&24.7&18.3&11.2&30.2 \\
			Mon 02-03 &10&&&24.1&17.7&11.2&29.7 \\
			Mon 03-04 &10&&&24.2&17.8&11.2&29.7 \\
			\bottomrule 
		\end{tabular}
		\caption{Hourly average computational results over 30 Mondays of simulated request data and average transportation data from Monday bus trips during first half of 2019.}
		\label{tab:30mondays}
	\end{threeparttable}
\end{table}

\begin{figure}[h]
	\begin{subfigure}[t]{0.5\textwidth}
		\centering\footnotesize
		\begin{tikzpicture}
			\begin{axis}[ybar, ymin = 0, xmin=0,xmax=8,width=1.1\textwidth, height=4.5cm, bar width = 0.2cm, tick pos=left,
				xticklabels={Mon, Tue, Wed, Thu, Fri, Sa, Su},xtick={1,2,3,4,5,6,7},
				legend style={at={(axis cs:0.5,45)},anchor=south west}]
				\addplot coordinates {
					(1,24.1) 
					(2,24.7) 
					(3,25.1) 
					(4,22.6) 
					(5,22.4) 
					(6,19.5) 
					(7,20.1) 
				};
				\addplot coordinates {
					(1,8.5) 
					(2,8.4) 
					(3,8.4) 
					(4,8.9) 
					(5,8.1) 
					(6,7.7) 
					(7,8.1)
				};
				%\legend {\emph{Bus service}, \emph{Ridepooling}};
			\end{axis}
		\end{tikzpicture}
		\caption{Average regret.} \label{fig:sub_regret}
	\end{subfigure}\hfill
	\begin{subfigure}[t]{0.5\textwidth}
		\centering\footnotesize
		\begin{tikzpicture}
			\begin{axis}[ybar, ymin = 0, xmin=0,xmax=8,width=1.1\textwidth, height=4.5cm, bar width = 0.2cm, tick pos=left,
				xticklabels={Mon, Tue, Wed, Thu, Fri, Sa, Su},xtick={1,2,3,4,5,6,7},
				legend style={at={(axis cs:0.5,45)},anchor=south west}]
				\addplot coordinates {
					(1,17.7) 
					(2,19.0) 
					(3,19.0) 
					(4,16.1) 
					(5,16.8) 
					(6,13.7) 
					(7,14.6) 
				};
				\addplot coordinates {
					(1,7.1) 
					(2,7.0) 
					(3,7.2) 
					(4,7.5) 
					(5,6.9) 
					(6,6.7) 
					(7,7.0)
				};
			\end{axis}
		\end{tikzpicture}
		\caption{Average waiting time.} \label{fig:sub_wait}
	\end{subfigure}
	\vspace*{0.5cm}
	\begin{subfigure}[t]{0.5\textwidth}
		\centering\footnotesize
		\begin{tikzpicture}
			\begin{axis}[ybar, ymin = 0, xmin=0,xmax=8,width=1.1\textwidth, height=4.5cm, bar width = 0.2cm, tick pos=left,
				xticklabels={Mon, Tue, Wed, Thu, Fri, Sa, Su},xtick={1,2,3,4,5,6,7},
				legend style={at={(axis cs:0.5,45)},anchor=south west}]
				\addplot coordinates {
					(1,11.2) 
					(2,10.7) 
					(3,11.0) 
					(4,11.3) 
					(5,10.5) 
					(6,10.5) 
					(7,10.2) 
				};			
				\addplot coordinates {
					(1,6.8) 
					(2,6.8) 
					(3,6.5) 
					(4,6.8) 
					(5,6.5) 
					(6,6.1) 
					(7,6.6)
				};
			\end{axis}
		\end{tikzpicture}
		\caption{Average ride time.} \label{fig:sub_ride}
	\end{subfigure}\hfill
	\begin{subfigure}[t]{0.5\textwidth}
		\centering\footnotesize
		\begin{tikzpicture}
			\begin{axis}[ybar, ymin = 0, xmin=0,xmax=8,width=1.1\textwidth, height=4.5cm, bar width = 0.2cm, tick pos=left,
				xticklabels={Mon, Tue, Wed, Thu, Fri, Sa, Su},xtick={1,2,3,4,5,6,7},
				legend style={at={(axis cs:0.5,45)},anchor=south west}]
				\addplot coordinates {
					(1,29.7) 
					(2,30.4) 
					(3,30.7) 
					(4,28.1) 
					(5,28.0) 
					(6,25.0) 
					(7,25.6) 
				};						
				\addplot coordinates {
					(1,14.7) 
					(2,14.6) 
					(3,14.5) 
					(4,15.1) 
					(5,14.1) 
					(6,13.6) 
					(7,14.4)
				};
			\end{axis}
		\end{tikzpicture}
		\caption{Average transportation time.} \label{fig:sub_trans}
	\end{subfigure}
	\caption{Comparison of dial-a-ride (red) and bus service (blue) for the case of scenario C.}
	\label{fig:C_comparison}
\end{figure}
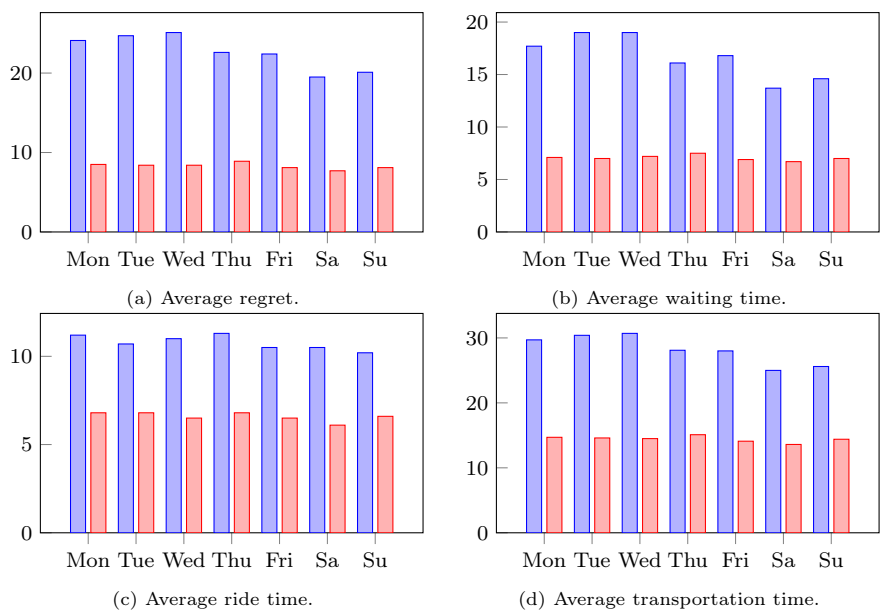

It becomes obvious from Table~\ref{tab:30weeks}, that we can achieve a much higher service quality by ridepooling than by the bus service. While the daily average regret ranges from 9--11 minutes in scenario A, 11--14 minutes in scenario B and 7--9 minutes in scenario C, using the bus service the average regret is between 19 and 26 minutes. Similar tendencies hold for average waiting time, average ride time and average transportation time. 
In scenario A, where the number of vehicles was selected so that, on average, eight transport requests during the hour of highest request density can be accepted, on average at most 1.2\% of the transport requests are denied. While this is an acceptable ratio for a dial-a-ride service, we note that the replacement of bus service generally requires a full coverage of all transport requests. 
Thus, also scenario B, where one vehicle less per weekday is used than in scenario A, is not eligible for a fair comparison. In scenario C, where one vehicle more than in scenario A is used, the average percentage of denied requests is 0.0\% for all weekdays but Monday, where it is 0.1\%. Hence, the size of the vehicle fleet for Mondays should be increased to seven vehicles, while for the remaining week the initial estimate is sufficient. Taking a closer look at scenario C, it becomes evident, that using the bus service to transport late evening requests, average regret and average waiting time are about 2.5 to 3 times as high, while average ride time and transportation time are about 1.5 to 2 times as high as compared to using the ridepooling service.
These ratios can be improved even further when taking into account the missing vehicle on Monday in scenario C:  we now take a look at the hourly results for Monday evening. Note, that the average results reported in Table~\ref{tab:30mondays}  are computed  individually for every hour, which explains for example the fact that the percentage of denied requests from 3 a.m.\ to 4 a.m.\ is higher than the percentage of denied requests for Mondays as a whole day. From Table~\ref{tab:30mondays} we deduce that the number of vehicles we assumed for Monday is sufficient, only for the hours of 3 a.m.\ to 4 a.m., an additional vehicle is needed. In this hour, also the average regret, waiting time, ride time and transportation time is larger than in the rest of the evening, so that the total averages for Monday are increased by this hour with too few vehicles. 
 A summary of the difference in service quality for the case of scenario C can be found in Figure~\ref{fig:C_comparison}. We can conclude that there is a high potential in the replacement of bus services by ridepooling during late evening hours if quantitative criteria are used to evaluate the service quality. As discussed previously, a vehicle fleet of 4--7 vehicles would be needed, depending on the weekday, to be able to serve all requests and to replace the 10 buses needed to serve all bus lines simultaneously. We note that we do not compare the service costs in this study. The strong improvement in the overall service quality opens the door to an even better and potentially more profitable service, as people might become more willing to use public transport instead of private cars. As ridepooling services generally benefit from a high demand, so that rides can be shared, the results of this study are indeed promising.

%\begin{figure}
%	\begin{subfigure}[t]{0.5\textwidth}
%		\centering\footnotesize
%		\includegraphics[scale = 0.4]{Visualizations/Avg_regret.png}
%		\caption{Average regret\label{fig:sub_regret}.}
%	\end{subfigure}\hfill
%	\begin{subfigure}[t]{0.5\textwidth}
%		\centering\footnotesize
%			\includegraphics[scale = 0.4]{Visualizations/Avg_wait.png}
%		\caption{Average waiting time\label{fig:sub_wait}.}
%	\end{subfigure}
%
%		\begin{subfigure}[t]{0.5\textwidth}
%		\centering\footnotesize
%		\includegraphics[scale = 0.4]{Visualizations/Avg_ride.png}
%		\caption{Average ride time\label{fig:sub_ride}.}
%	\end{subfigure}\hfill
%	\begin{subfigure}[t]{0.5\textwidth}
%		\centering\footnotesize
%		\includegraphics[scale = 0.4]{Visualizations/Avg_transportation.png}
%		\caption{Average transportation time\label{fig:sub_trans}.}
%	\end{subfigure}
%	\caption{Comparison of the service quality of ridepooling (red) and bus service (blue) for the case of scenario C. }
%	\label{fig:scenarioC}
%\end{figure}

\section{Conclusions}\label{sec:concl}
In this paper we investigate the effects on the service quality when replacing buses during late evening hours by a ridepooling service. This is motivated by the observation that particularly late at night, the number of travel requests is usually rather small so that ridepooling services could offer a good alternative to regular bus services.
Moreover, line based bus systems are not working to capacity during these hours. For the purpose of this investigation 30 weeks of simulated transport requests were created from bus trip and ridepooling data sets in the city of  Wuppertal in Germany by means of a predicitve simulation. The transportation of the simulated requests using the ridepooling service was mimicked by a rolling-horizon algorithm based on the iterative solution of MILPs. This algorithm was enhanced by a feasible-path heuristic to cope with a high request density. The computational results show that there is a large potential for improvement of the service quality when offering ridepooling services as compared to bus services: Using the ridepooling service, the average transportation time can be reduced by about 50\%. One shortcoming of our study is, that due to lack of data on parts of the bus service, we were not able to compare the total vehicle kilometers driven. Although one could argue that the total vehicle kilometers driven are not a primary performance indicator due to the fact the ridepooling cabs are often electric vehicles, it would still be interesting to compare them to better assess the cost of an improved service quality. Also, from WSW statistics we know that the number of actual passengers is about 3 times as high as the number of WSW-LAN users.
In the future, this computational study based on historic and simulated data-sets should be complemented by customer surveys investigating the acceptance of ridepooling as a supplement of line based bus services. Besides the service quality and environmental aspects of the transportation system, practical considerations on a low-threshold user-interface and communication are important to provide good user experience also for late-adopters. Moreover, in the future we plan to adapt the MILP model to handle variable sizes of the vehicle fleet depending on the request density per hour.

\section*{Acknowledgements}
\noindent This work was partially supported by the state of North Rhine-Westphalia (Germany) within the project ``bergisch.smart.mobility''. We are thankful to WSW and \hma for providing data for this study.

\newpage
\bibliographystyle{abbrvnat}
\bibliography{main_ridehailing}

\newpage
\appendix
%\counterwithout{table}{section}

\section{Parameters}\label{sec:parasvars}

\begin{table}[h]
	\centering
	\begin{threeparttable}
		\small
		\begin{tabularx}{\textwidth}{cX}
			\toprule
			Parameter & Description \\
			\midrule
			$n$ & number of  transport requests \\
			$\Delta$ & time allowed to communicate answer to requests \\
			$\tau_i - \Delta$ & time at which request $i$ is revealed \\
			$i^+$, $i^-$ & pick-up and drop-off location of request $i$ \\
			$0$ & vehicle depot \\
			$Q$ & vehicle capacity \\
			$s_i$ & service duration associated with request $i$ \\
			$\left[e_j, \ell_j \right]$ & time window associated with location $j$ \\
			$c_{j_1j_2}$, $t_{j_1j_2}$ &  routing cost and travel time on arc $a$ \\
			$G(\tau_j)$ & dynamic event-based graph corresponding to time $\tau_j$ \\
			$\rho_i$ & number of feasible paths associated with request $i$ \\
			$\rho_{\text{abs}}$, $\rho_{\text{rel}}$ & minimum absolute number and mininum percentage of feasible paths allowed \\
			$B_i$ & beginning of service at pick-up location request $i$ \\
			$D_i$ & departure time at pick-up location request $i$\\
			$A_i$ & arrival time at drop-off location request $i$ \\
			$p^\star$ & binary vector representing acceptance or denial of a request \\ 
			\bottomrule
			
		\end{tabularx}
		
		\caption{List of parameters.}
		\label{paras}
	\end{threeparttable}
\end{table}

\end{document}